\documentclass[conference]{IEEEtran}
\IEEEoverridecommandlockouts
\usepackage{cite}
\usepackage{hyperref}
\usepackage{amsmath,amssymb,amsfonts}
\usepackage{algorithmic}
\usepackage{graphicx}
\usepackage{textcomp}
\usepackage{xcolor}
\def\BibTeX{{\rm B\kern-.05em{\sc i\kern-.025em b}\kern-.08em
    T\kern-.1667em\lower.7ex\hbox{E}\kern-.125emX}}
\begin{document}

\title{PrivacyCube: A Tangible Device for Improving Privacy Awareness in IoT\\
}
\author{
    Bayan Al Muhander\\
    \textit{Cardiff University}\\
    Cardiff, UK\\
   almuhanderb@cardiff.ac.uk
  \and
    Omer Rana\\
   \textit{Cardiff University}\\
    Cardiff, UK\\
   RanaOF@cardiff.ac.uk
    \and
    Nalin Arachchilage\\
    \textit{University of Auckland}\\
    Auckland, New Zealand \\
    nalin.arachchilage@auckland.ac.nz
    \and
    Charith Perera\\
   \textit{Cardiff University}\\
    Cardiff, UK\\
    pererac@cardiff.ac.uk
}



\maketitle

\begin{abstract}
Consumers increasingly bring IoT devices into their living spaces without understanding how their data is
collected, processed, and used. We present PrivacyCube, a novel tangible device designed to explore the extent to which privacy awareness in smart homes can be elevated. PrivacyCube visualises IoT
devices’ data consumption displaying privacy-related notices. PrivacyCube aims at assisting families to (i) understand key privacy aspects better and (ii) have conversations around data management practices of IoT devices. Thus, families can learn and make informed privacy decisions collectively.
\end{abstract}


\section{Introduction}
Personal data protection regulations, such as the General Data Protection Regulation (GDPR) and the California Consumer Privacy Act (CCPA), have enforced transparency while processing consumers' data. Data controllers must provide individuals with privacy notices about existing or potential data processing practices. Prior studies discussed that effective privacy notices support individuals in making informed privacy decisions \cite{acquisti2017nudges}. However, available notices are largely ignored or abandoned and forgotten over time. Studies have linked the failure of the privacy notices to the way they are presented, where most of them are often long and difficult to read \cite{schaub2017designing}.  

\begin{figure}[t]
\centering
\includegraphics [scale=.35]{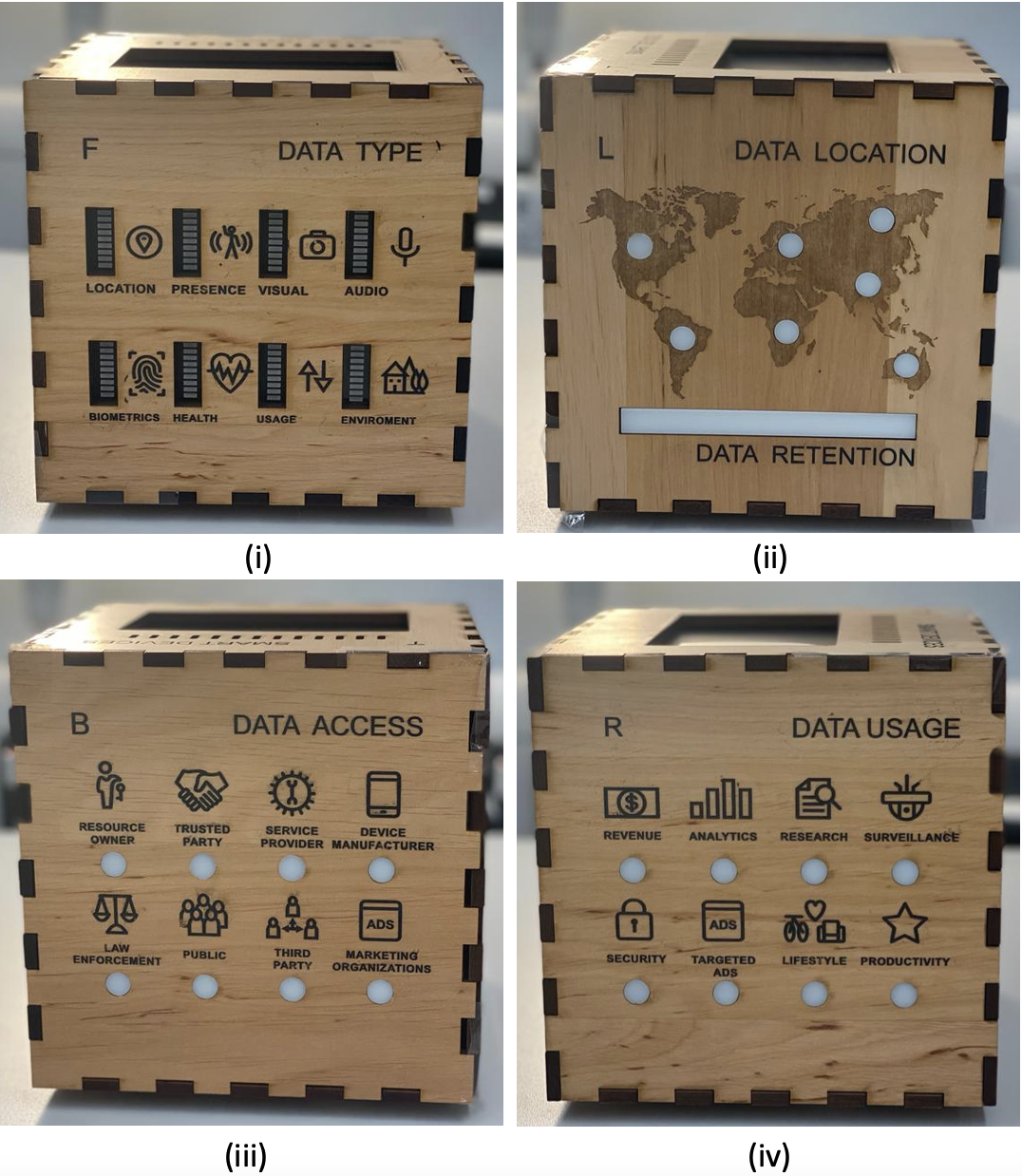}

\caption{PrivacyCube four faces: (i) collected data, (ii) data storage location and retention period, (iii) data access, and (iv) data usage. (\href{https://iotgarage.net/projects/demos/BayanIOTDI2022Demo.html}{Demo Video})}. 

\label{fig:11} 

\end{figure}

\begin{figure*}[t!]
\centering

\includegraphics [scale=.185]{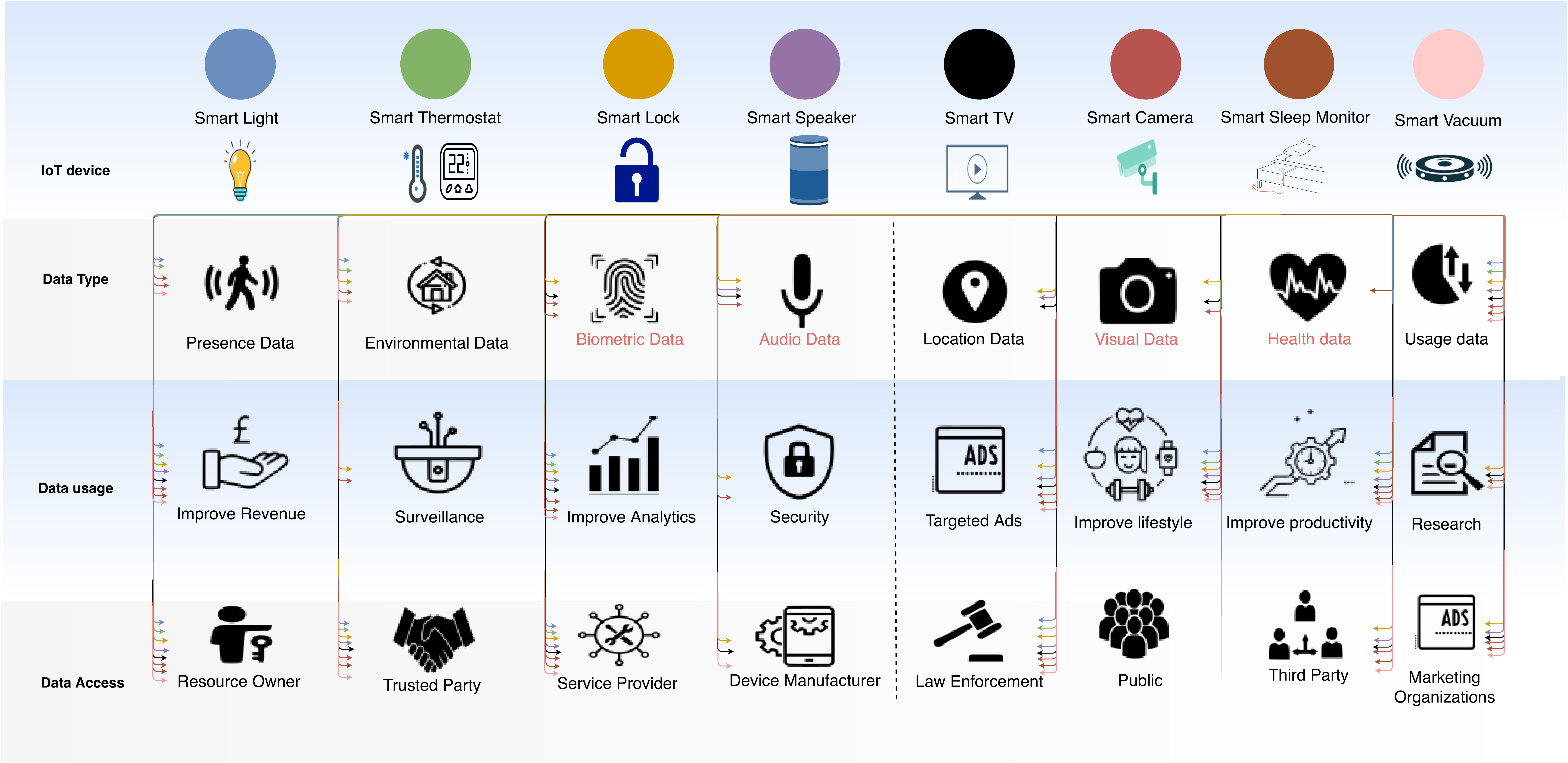}

\caption{Mapping IoT devices on PrivacyCube to indicate data type, use, and access. Each IoT device has a unique colour that is indicated at the top of the device icon. Identifiable data is shown in red, while non-identifiable data is shown in black.} \label{fig:1}

\end{figure*}

Existing privacy notices designs are often limited by the information presented to the data subjects, or are websites specific and not IoT related \cite{korayem2016enhancing}. Moreover, many of the available privacy notices are static in nature, meaning they do not receive updates and only display one type of notice with no variation based on device/data type and use. The designed tangible device in this paper is set to reform the presentation of the privacy notices to include more meaningful information for the user. Physical cubes, unlike mobile apps, have long been used as a learning and exploration tool because they are an appealing embodiment of more complicated concepts \cite{ishii1997tangible,houben2016physikit}. PrivacyCube can provide variable privacy notices based on the used IoT device and the data it  collects \cite{muhander2020privacy}.

In this work, we prototype PrivacyCube, a physical tangible notices for improving privacy awareness in smart homes, refer to Figure \ref{fig:11}. PrivacyCube will provide users with necessary data practices of the IoT resources in the vicinity in an interactive way. The notices displayed on PrivacyCube are largely based on a privacy infrastructure developed by Das et al. \cite{das2018personalized}. PrivacyCube will display the active IoT resources and emit light for each data usage practice.

\section{IMPLEMENTATION}

PrivacyCube is an interactive device that allows users to explore and learn about the surrounding IoT devices and the main data practises that the device performs on their data. It is composed of a wooden casing shaped like a cube. The casing is laser cut from 5 mm thick plywood and can be easily assembled by slotting together with no tools required. There is a 5-inch touch screen displaying the IoT devices on the top face. The bottom face serves as the cube's base, while the other faces show data-related privacy notices. PrivceyCube will emit lights as interactive privacy notices to users to mimic actual data collection activity. Users can use the touch screen to interact with PrivacyCube and explore related privacy notices.

 Individuals are interested in receiving notifications related to the collection, usage, storage of their data \cite{das2018personalized}. Therefore, the data practices included in PrivacyCube notify individuals about the collected data type, data usage, data access, data location, and data retention period. Based on multiple privacy policies templates included in the IoT Assistant App \cite{feng2021design}, we created a graph, Figure \ref{fig:1}, mapping the IoT devices displayed on PrivacyCube. Each cube face represents different data practices related to IoT as pictured in Figure \ref{fig:11}. In the event of a data collection, the top face will show the active IoT device, while the other faces will show the associated data practices.

\section{DEMONSTRATION}
We evaluated PrivacyCube against the notion of \textit {“meaningful privacy choices” }\cite{feng2021design}. Because PrivacyCube only provides visualisation and not the configuration of data use practices, our evaluation includes three of the five facets (i.e. user awareness, comprehensiveness, and neutrality). PrivacyCube simplifies learning about active IoT devices and their data usage practices. This feature can increase user awareness of nearby IoT devices and their privacy considerations. Integrating the identification of IoT devices and their privacy notices in a single tangible design delivers comprehensive information regarding IoT data practices to assist users in understanding the implications of their privacy decisions. In addition, we believe that the cube design provides a neutral ground for home occupants to engage with their privacy notifications.

The following use case, about an individual named Alex, demonstrates PrivacyCube functionality and how it communicates information to raise individuals' privacy awareness. Alex protects his home with a smart lock that includes a built-in camera. Alex finds it convenient to recognise and communicate with visitors to his home via his smart device. Alex does not understand that for the smart lock to function; it must detect and store sensitive data that can be used for purposes other than home security. The deployment of PrivacyCube will inform Alex about what happens to his data in the following:

\begin{itemize}
    \item The lock icon on the top face of PrivacyCube will light up indicating the IoT device in-use.
 \item The environmental, biometric, audio, location, visual, and usage icons on the front face will light up indicating the type of data the device is collecting. 
 \item The time-bar and world map will light up some sections indicating where and for how long the data is stored.
 \item The revenue, surveillance, analytics, security, targeted ads, lifestyle, productivity, and research icons will light up indicating the data usages.
 \item The resource owner, trusted party, service provider, device manufacturer, law enforcement, third party, and marketing organizations will light up indicating data access. 
\end{itemize}


\bibliographystyle{unsrt}
\bibliography{sample-base}

\begin{thebibliography}{1}

\bibitem{acquisti2017nudges}
Alessandro Acquisti, Idris Adjerid, Rebecca Balebako, Laura Brandimarte,
  Lorrie~Faith Cranor, Saranga Komanduri, Pedro~Giovanni Leon, Norman Sadeh,
  Florian Schaub, Manya Sleeper, et~al.
\newblock Nudges for privacy and security: Understanding and assisting users’
  choices online.
\newblock {\em ACM Computing Surveys (CSUR)}, 50(3):1--41, 2017.

\bibitem{schaub2017designing}
Florian Schaub, Rebecca Balebako, and Lorrie~Faith Cranor.
\newblock Designing effective privacy notices and controls.
\newblock {\em IEEE Internet Computing}, 2017.

\bibitem{korayem2016enhancing}
Mohammed Korayem, Robert Templeman, Dennis Chen, David Crandall, and Apu
  Kapadia.
\newblock Enhancing lifelogging privacy by detecting screens.
\newblock In {\em Proceedings of the 2016 CHI Conference on Human Factors in
  Computing Systems}, pages 4309--4314, 2016.

\bibitem{ishii1997tangible}
Hiroshi Ishii and Brygg Ullmer.
\newblock Tangible bits: towards seamless interfaces between people, bits and
  atoms.
\newblock In {\em Proceedings of the ACM SIGCHI Conference on Human factors in
  computing systems}, 1997.

\bibitem{houben2016physikit}
Steven Houben, Connie Golsteijn, Sarah Gallacher, Rose Johnson, Saskia Bakker,
  Nicolai Marquardt, Licia Capra, and Yvonne Rogers.
\newblock Physikit: Data engagement through physical ambient visualizations in
  the home.
\newblock In {\em Proceedings of the 2016 CHI Conference on Human Factors in
  Computing Systems}, pages 1608--1619, 2016.

\bibitem{muhander2020privacy}
Bayan AlMuhander, Jason Wiese, Omer Rana, and Charith Perera.
\newblock Privacy-aware internet of things notices in shared spaces: A survey.
\newblock {\em arXiv preprint arXiv:2006.13633}, 2020.

\bibitem{das2018personalized}
Anupam Das, Martin Degeling, Daniel Smullen, and Norman Sadeh.
\newblock Personalized privacy assistants for the internet of things: Providing
  users with notice and choice.
\newblock {\em IEEE Pervasive Computing}, 2018.

\bibitem{feng2021design}
Yuanyuan Feng, Yaxing Yao, and Norman Sadeh.
\newblock A design space for privacy choices: Towards meaningful privacy
  control in the internet of things.
\newblock In {\em Proceedings of the 2021 CHI Conference on Human Factors in
  Computing Systems}, pages 1--16, 2021.

\end{thebibliography}

\end{document}